\documentclass[twocolumn,secnumarabic,amssymb,
amsmath,nobibnotes, aps, showpacs,prl]{revtex4}

\usepackage{graphicx}
\usepackage{bm}

\def\be{\begin{equation}}
\def\ee{\end{equation}}

\begin{document}
                                                                                
\title{CMB Polarization Systematics Due to Beam Asymmetry:\\
Impact on Cosmological Birefringence}

\ \\
\author{N.J. Miller, M. Shimon, B.G. Keating}
\affiliation{Center for Astrophysics and Space Sciences, University
of California, San Diego, 9500 Gilman Drive, La Jolla, CA, 92093-0424}
                                                                
\date{\today}

                                                                              
\begin{abstract}
The standard cosmological model is assumed to respect parity symmetry. 
Under this assumption the cross-correlations of the CMB's temperature 
anisotropy and `gradient'-like polarization, with the `curl'-like 
polarization identically vanish over the full sky. However, extensions 
of the standard model which allow for light scalar field or axion coupling 
to the electromagnetic field, or coupling to the Riemann gravitational 
field-strength, as well as other modifications of field theories, may 
induce a rotation of the CMB polarization plane on cosmological scales 
and manifest itself as nonvanishing TB and EB cross-correlations. 
Recently, the degree of parity violation (reflected in polarization rotation) 
was constrained using data from BOOMERANG, WMAP and QUAD. Forecasts have been made 
for near-future experiments (e.g. PLANCK) to further constrain parity- 
and Lorentz-violating terms in the fundamental interactions of nature. 
Here we consider a real-world effect induced by a class of telescope 
beam systematics which can mimic the rotation of polarization plane 
or otherwise induce nonvanishing TB and EB correlations. 
In particular, adopting the viewpoint that the primary target of future 
experiments will be the inflationary B-mode signal, we assume the 
beam-systematics of the upcoming PLANCK and POLARBEAR experiments are 
optimized towards this goal, and explore the implications of the allowed 
levels of beam systematics on the resulting precision of polarization-rotation 
measurements.
\end{abstract}

\maketitle
                                                                                
\section{Introduction}

Future CMB polarimeters are poised to detect the B-mode polarization. 
It is expected that PLANCK will 
detect the lensing-induced B-mode in the near future and a host of CMB 
experiments are expected to detect the horizon-scale inflationary B-mode 
signal if the tensor-to-scalar ratio is $r\approx 0.01$. This extremely 
feeble signal is susceptible to a variety of foregrounds 
and systematics which may severely degrade the science which CMB 
observations could otherwise do in an ideal-world. Analytic expressions 
have been obtained for several types of beam 
systematics as well as the acceptable levels of beam uncertainties, such 
as differential ellipticity, gain, rotation, etc. (e.g. [1]-[4]). 
We are especially concerned with those beam features whose levels and other 
characteristics are not precisely known (e.g., the extent to which the 
beam is elliptical, its gain mismatch, etc) since this cannot be remedied 
in the process of data analysis and will therefore add an uncertainty 
to the CMB power spectra. The level of allowed beam uncertainty is determined 
mainly by the type of science one could do with B-mode polarization; 
these are associated with inflation and lensing-related science such 
as neutrino masses, dark energy equation of state, etc. 
It might be possible that an actual beam calibration will result in 
a better calibration than the minimum requirements for the science.
As it turns out, for as low a tensor-to-scalar ratio as $r=0.01$ 
the most stringent experimental requirements on the allowed level 
of beam uncertainty come from the inflationary science. 
Not only will B-mode observations open a unique window to inflation 
and gravitational lensing, it also offers a rare opportunity to peer 
into two of the most fundamental underlying symmetries of physics, 
namely - parity and Lorentz invariance. With an horizon-scale optical 
path we have the longest possible leverage-arm to detect the accumulated 
effect of `cosmic birefringence' (CB), a continuous 
rotation of the polarization plane of the electromagnetic radiation. 
Lorentz- and/or parity-violating terms in the electromagnetic or 
gravitational sectors of a hypothetical unified model of the fundamental 
interactions of nature may induce such a CB (e.g. [5]-[10]). 
Searching evidence for CB in the CMB power spectra have been the 
objective of several recent works ([11]-[16]) which analyzed the data 
in an ideal, systematics-free, world.

The purpose of this work is to highlight the importance of controlling 
beam systematics, mainly their effect on the polarization state of the 
CMB and in the context of CB. A tantalizing idea is to  
use the TB and EB cross-correlations in the CMB to detect Lorentz-symmetry 
breaking or parity-odd features. 
However, these cross-correlations are especially susceptible to even 
the smallest degradation in the B-mode (due to temperature leakage 
or beam rotation) and therefore a careful study of the implications 
beam systematics may have on the CB science is well motivated.

The outline of this paper is as follows.
Section 2 summarizes CMB power spectra and relevant Fisher matrix relations. 
The effect of beam systematics is discussed in section 3. 
We describe our analysis and results in section 4 followed by a 
discussion in section 5. We conclude in section 6.
Our basic calculations are supplemented with two appendices: 
In Appendix A we estimate the resulting level of beam ellipticity 
assuming a standard procedure of telescope beam calibration 
(which determines the level of TB and EB correlations induced 
by beam ellipticity). Appendix B is dedicated to a brief discussion 
of the randomness of beam-pair orientation in the 
focal plane and its implications on the spurious power spectra.

\section{CMB Power Spectra and CB}

Before we describe the effect of non-standard physics on the CMB, let us first review 
the basics of the CMB statistics in the {\it standard} cosmological model because 
detecting this physics relies on deviations from the standard statistics of the CMB.
The temperature $T$ and the Stokes parameters $Q$ \& $U$ 
are expanded in scalar and rank $\pm 2$ tensor spherical harmonics
\begin{eqnarray}
T(\hat{n})&=&\sum_{l,m}a^{T}_{lm}Y_{lm}(\hat{n})\nonumber\\
(Q\pm iU)(\hat{n})&=&\sum_{l,m}a_{\pm 2,lm}\ _{\pm 2}
Y_{lm}(\hat{n})
\end{eqnarray} 
where the $E$ and $B$ (scalar and pseudo-scalar, respectively) modes of polarization 
have expansion coefficients $E_{lm}$ and $B_{lm}$ which can be expressed in terms of 
the spin $\pm 2$ coefficients
\begin{eqnarray}
a_{\pm 2,lm}=a_{lm}^{E}\pm a_{lm}^{B}.   
\end{eqnarray}
It is assumed, and well motivated by cosmological inflation, that the temperature 
anisotropy and polarization modes can be drawn from gaussian distributions fully 
characterized by vanishing means and standard deviations described by the power spectra.
This implies that the ensemble averages of the expansion coefficients vanish
\begin{eqnarray}
\langle a^{X}_{lm}\rangle=0
\end{eqnarray}
where $X$ can assume either $T$, $E$ or $B$
and that the widths of the gaussians satisfy (by the assumption of statistical 
isotropy which is the multipole-space dual of real space isotropy)
\begin{eqnarray}
\langle a^{X*}_{l'm'}a^{X'}_{lm}\rangle=C_{l}^{XX'}\delta_{ll'}\delta_{mm'} 
\end{eqnarray}
and
\begin{eqnarray}
C_{l}^{XX'}\equiv\frac{1}{2l+1}\sum_{m=-l}^{l}a^{X*}_{lm}a^{X'}_{lm}.
\end{eqnarray}
In the standard model $E$ and $B$ are pure parity states (even and odd, respectively). 
The correlation, over the full sky, of the B-mode with either the temperature or 
E-mode polarization vanishes in this case. 
However, if the polarization plane is rotated for some reason 
(cosmological or otherwise), there will be a E-B mixing which will induce the `forbidden' TB 
and EB power spectra (leakage from the underlying TE and EE cross-correlations, respectively). 
Let us see how this effect takes place.
While rotation of polarization leaves the temperature invariant, the polarization 
state will be affected. Consider for example the flat- sky approximation, which 
holds for small angular scales only (and is fully applicable on beam scales), 
but enables a transparent derivation of the 
following relations between the original and observed degraded power spectra.
In this approximation the Stokes Q and U parameters, which fully describe the 
polarization state of a linearly polarized radiation, can be expanded in  
Fourier space as follows
\begin{eqnarray}
(Q\pm iU)({\bf x})=\frac{1}{2\pi}\int(E_{{\bf l}}\pm iB_{{\bf l}})
e^{i{\bf l}\cdot{\bf x}}e^{\pm 2i(\phi_{l}-\phi_{x})}d^{2}{\bf l}.
\end{eqnarray}
$Q\pm iU$ are pure spin $\pm 2$ states and under a rotation by $-\gamma$ 
pick up phases $e^{\pm 2i\gamma}$. As seen from Eq.(6) this results in an E-B mixing
\begin{eqnarray}
E'&=&E\cos 2\gamma-B\sin 2\gamma\nonumber\\
B'&=&E\sin 2\gamma+B\cos 2\gamma
\end{eqnarray}
and correlating with $T'=T$ we obtain the following power spectra
\begin{eqnarray}
C'^{TE}_{l}&=&C^{TE}_{l}\cos 2\gamma\nonumber\\
C'^{TB}_{l}&=&C^{TE}_{l}\sin 2\gamma\nonumber\\
C'^{EE}_{l}&=&C^{EE}_{l}\cos^{2}2\gamma+C^{BB}_{l}\sin^{2}2\gamma\nonumber\\
C'^{BB}_{l}&=&C^{EE}_{l}\sin^{2}2\gamma+C^{BB}_{l}\cos^{2}2\gamma\nonumber\\
C'^{EB}_{l}&=&\frac{1}{2}(C^{EE}_{l}-C^{BB}_{l})\sin 4\gamma.
\end{eqnarray}
Note that no assumption has been made here as to the origin of this rotation, 
namely whether or not it is 
of cosmological origin. In the literature $\gamma$ is identified with the 
CB rotation angle (see e.g. [15] and [16]). We will see in the next section 
that, in practice, 
one has to account for other, non-cosmological effects which mimic cosmological 
polarization-rotation and may bias the results for the best-fit $\gamma$ if not 
properly accounted for (or, to a lesser extent, change the uncertainty of the 
inferred rotation itself). 
The main feature of Eq.(8) is that the `forbidden' $C^{TB}$ and $C^{EB}$ do not 
vanish in general, and it is this property of the power spectra which will be used 
(in the near-future, once we have high signal-to-noise measurements of TB and EB 
correlations) to 
infer the level of CB. Current weak upper limits on $\gamma$ are derived from 
comparing $C_{l}^{TT}$ with $C_{l}^{TE}$ and $C_{l}^{EE}$, all of which are 
primarily sourced by density perturbations.
In the absence of CB, $C_{l}^{TT}$ directly fixes $C_{l}^{TE}$ and $C_{l}^{EE}$ 
and the latter add little new cosmological information in the standard model 
with no reionization. Any deficit in the latter is ascribed to E-B rotation.

\section{The Effect of Beam Systematics}

No CMB observation is perfect and is always confronted, at some level, by optical 
beam systematics. Beam mismatch can induce temperature leakage to polarization as 
well as mode-mixing between the E- and B-modes ([1], [17]). Temperature leakage to 
polarization is considered the most pernicious systematic due to the fact that 
the primordial temperature anisotropy is at least an order of magnitude larger 
than the E-mode and two-three orders of magnitude larger than the B-mode. 
Therefore, even a small level of temperature leakage to polarization significantly 
contaminates the signal. 
For CB science, however, we are most concerned with cross-polarization and we open 
this discussion with the effect of beam rotation which mixes E and B-modes ([1], [17]). 
Beam rotation by an angle $\varepsilon$ trivially biases $\gamma$ at this level which 
implies that the uncertainty on the inferred $\gamma$ cannot be smaller than the bias 
induced by the uncertainty in beam orientation, $\varepsilon$. Threshold values for 
$\varepsilon$ have been found for POLARBEAR, PLANCK and CMBPOL [3] based on the 
requirement that both inflation and lensing science are not degraded above 
certain level. However, this requirement does not necessarily protect the 
CB science from beam systematics.

CMB polarimeters such as PLANCK will have sufficiently high instrumental sensitivity 
to allow a statistically significant detection of $C^{TB}$ (in a systematics-free world), 
the `smoking gun' for CB. 
As will be shown below, PLANCK and POLARBEAR have the raw sensitivity to detect 
rotation angles as small as few arcminutes. Any beam rotation by a larger angle 
will wash out the cosmological 
rotation of the polarization plane if CB is small enough. In these high-sensitivity experiments high 
polarimetric fidelity is required.

In the following we briefly discuss what impact two other beam systematics have on CB. 
These two effects, due to beam ellipticity and pointing error, leak temperature to 
polarization and are not a simple mixing of polarization as is the case with beam rotation. 
Since these two offsets will have a unique spectral-dependence in Fourier space 
(the systematic B-mode power spectra due to differential pointing and ellipticity 
scale as $\propto l^{2}C_{l}^{T}$ and $\propto l^{4}C_{l}^{T}$, respectively) they 
can be {\it statistically} distinguished and separated (with some residual `noise'), 
in principle, from CB which inherits the 
spectral shape of $C_{l}^{TE}$ 
Eqs.(7) will not be preserved in this case and we do expect that the likelihood 
function will change its {\it shape} and not be only biased. Table I summarizes 
the effect of differential rotation, pointing and ellipticity studied in [17] 
on $C_{l}^{TB}$ and $C_{l}^{EB}$.
It should be mentioned here that two other beam systematics considered in [17] and [3], 
differential gain and beamwidth, do not contribute to $C_{l}^{TB}$ and $C_{l}^{EB}$. 
They do, however, affect $C_{l}^{TE}$, $C_{l}^{EE}$ and $C_{l}^{BB}$, but since our 
main goal here is to test the susceptibility of $C_{l}^{TB}$ and $C_{l}^{EB}$ to beam 
systematics and the implications it has on the inferred $\gamma$ we do not address 
the effects of differential gain and beamwidth in this work.

\section{Analysis and Results}

An easy way to assess parameter uncertainty ($\gamma$, the CB rotation angle in our case) 
is to invoke the Fisher information-matrix approach (though not in the presence of 
significant bias, as discussed below).
The elements of the Fisher matrix are defined as follows
\begin{eqnarray}
F_{ij}=\frac{1}{2}\sum_{l}(2l+1)f_{sky}{\rm Trace}
[{\bf C}^{-1}\frac{\partial{\bf C}}{\partial\lambda_{i}}
{\bf C}^{-1}\frac{\partial{\bf C}}{\partial\lambda_{j}}]
\end{eqnarray}
where the 4-D symmetric covariance matrix ${\bf C}_{l}$ (per each multipole $l$) is defined as 
\begin{eqnarray}
	{\bf C}_{l} &\equiv& \left(\begin{array}{c c c c}
C_{l}^{TT} & C_{l}^{TE} & C_{l}^{TB} & C_{l}^{Td}\\
C_{l}^{TE} & C_{l}^{EE} & C_{l}^{EB} & 0\\
C_{l}^{TB} & C_{l}^{EB} & C_{l}^{BB} & 0\\
C_{l}^{Td} & 0 & 0 & C_{l}^{dd}
\end{array}\right)
\end{eqnarray}
where $C_{l}^{dd}$ and $C_{l}^{Td}$ refer to the power spectra associated with the lensing deflection angle [3].
The partial derivatives are taken with respect to the parameters $\lambda_{i}$. 
Here we allow nonvanishing $TB$ and $EB$ correlations and include lensing extraction.
The dimensionality of the Fisher matrix is $N\times N$ where $N$ is the dimension of 
parameter space of the cosmological model in question. The $1\sigma$ error on the 
parameter $\lambda_{i}$ is given by
\begin{eqnarray}
\sigma_{\lambda_{i}}=\sqrt{({\bf F}^{-1})_{ii}}.   
\end{eqnarray} 
The 1$\sigma$ error obtained by this procedure only sets a lower limit on the actual 
error. In general, when 
parameter degeneracy is high, as well as in the presence of significant bias, one 
resorts to Monte Carlo Markov Chain 
(MCMC) analysis [3]. 
The terms $C_{l}^{TB}$ and $C_{l}^{EB}$ drop from the matrix ${\bf C}$ (Eq. 10) 
in the standard model, 
but in general, especially here, they do not. 

A small bias in a parameter $\lambda_i$ compared to the statistical uncertainty, 
is given by [2,3]
\begin{eqnarray}
\Delta\lambda_{i} = \langle \lambda_i^{obs} \rangle - \langle \lambda_i^{true} 
\rangle = \sum_j \left( \mathbf{F}^{-1} \right) _{ij} B_j
\end{eqnarray}
where the bias vector $\mathbf{B}$ can be written as
\begin{eqnarray}
	\mathbf{B} = \sum_l (\mathbf{C}_{l}^{sys})^t \mathbf{\Phi}^{-1} 
\frac{\partial \mathbf{C}_{l}^{cmb}}{\partial \lambda_j},
\end{eqnarray}
$\mathbf{\Phi}_{ij}\equiv\textnormal{cov}(C_l^i,C_l^j)$ and $\mathbf{C}_l$ 
is a vector containing all six power spectra.
Our analysis includes lensing reconstruction and is essentially a generalization 
of [3].

As pointed above already, the underlying assumption of this work is that the 
primary target of upcoming CMB experiments is the inflationary B-mode; 
lensing-induced B-mode as well as the B-mode polarization induced by the 
Chern-Simons-type interaction term are only secondary. In addition, it 
was shown already [3] that the inflationary signal is more prone to beam 
systematics than the lensing signal is and therefore it was assumed that 
optimizing the beam systematics to the former should be sufficient for 
the latter.
Here, we use the analytic expressions for beam systematics found in [17] 
to determine the resulting bias in the inferred CB for both PLANCK and 
POLARBEAR. 
Before presenting the results of our Fisher-matrix analysis it is constructive 
to plot the systematic $C_{l}^{TB}$ and $C_{l}^{EB}$. These are shown for both PLANCK 
and POLARBEAR in Figure 1. Black (dotted), blue (dashed) and yellow (dotted-dashed) curves correspond to the systematics 
induced by differential pointing, ellipticity and beam-rotation, respectively (note 
that the effect of pointing on the EB cross-correlation is negligible and falls below 
the scale shown). For all the effects we assumed worst-case scanning strategy and 
polarimeter orientation as will be discussed below and extensively discussed in [17] and [3].
Red (continuous) curves show levels of $C_{l}^{TB}$ and $C_{l}^{EB}$ which these experiments could have 
detected (obtained from the Fisher-matrix analysis) in the absence of systematics. 
We refer to them as `nominal CB' detection. 
For comparison, the green (continuous) curve corresponds to a much larger rotation of the polarization 
plane that will result in $C_{l}^{TB}$ and $C_{l}^{EB}$ roughly three times larger than the 
largest systematics-induced TB and EB correlations. It is evident from the figure that 
realizing the nominal potential of PLANCK and POLARBEAR to detect CB via the TB and EB correlations 
will require a much better control of systematics than the level set by the inflationary-induced 
B-mode detection requirements, for example.

The results of our Fisher-matrix analysis are shown in Table II. Nominal uncertainties in $\gamma$ 
(i.e. the statistical error in a systematics-free experiment) are compared 
with the bias induced by differential pointing and ellipticity, $\Delta\gamma$.
The values for the differential pointing $\rho$ and ellipticity $e$ are adopted from 
[3], consistent with the assumption that PLANCK and POLARBEAR are optimized for 
inflation and lensing science. In the analysis we used all information from 
multipoles $l\leq 1200$. Figure 1 suggests that the bias decreases if $l_{max}$ is smaller 
but this of course comes on the expense of statistical uncertainty. We repeated the analysis with 
lower $l_{max}$ which indeed resulted in lower bias but even for $l_{max}$ as small as 200 
it was unacceptably large. The effect of beam rotation results in a pure bias 
in the inferred $\gamma$ and is therefore trivial and not included in the table.
Again, we find that beam calibration levels that suffice for the 
inflationary and lensing science are unfortunately insufficient for 
the CB science and this is the main conclusion of this paper. Especially 
worrisome in this context are beam rotation and differential ellipticity. 
Both systematics are independent of the scanning strategy and therefore 
cannot be easily mitigated. In addition, beam rotation induces both EB 
and TB correlations which have the same spectral shape in multipole-space 
as the CB; one cannot distinguish a beam rotation from CB. While it is clear from 
our analysis that the allowed ellipticity based on the inflation+lensing 
requirements may not be sufficient for CB purposes, beam calibration in 
a real experiment will result in a better control of ellipticity than 
the minimum allowed by inflation/lensing. We show in Appendix A that both PLANCK 
and POLARBEAR will benefit from a reasonable control of ellipticity which 
will allow beam ellipticities lower than the minimum required for inflationary 
and lensing science and will therefore result in an essentially unbiased CB 
detection from the E-B correlations of the CMB 
in case the nominal values of the respective $\gamma$ are considered.
However, TB is still prone, even in this ideal case, to beam ellipticity. 
Beam rotation remains the paramount concern, and of all types of systematics 
we identify it as the main obstacle towards CB detection; the nominal 
$\gamma$ values are typically a factor $\approx 10$ smaller than the allowed 
beam rotation (Table 2). Recently, it was shown how optimal 
estimators, which can filter spatially-dependent rotation, can be used to 
filter-out a spatially-dependent rotation due to the non-standard statistics 
it induces by inducing $l-l$ mode coupling [18]. However, the {\it constant} beam 
rotation we consider here cannot be distinguished from CB by this method.

The systematics induced by differential pointing depend, to first order, 
on the scanning strategy and for a non-ideal, yet uniform, scanning strategy 
we encapsulate the relevant information in the two scanning functions $f_{2}$ 
and $f_{3}$ ([17]) whose exact definitions are irrelevant to this discussion 
since we always assume the worst case scenario ([3]) $f_{i}=2\pi$.

The cosmological model adopted in our numerical calculations closely follows 
the WMAP5 results [15]; 
the baryon, cold dark mater, 
and neutrino physical energy densities in critical density units $\Omega_{b}h^{2}=0.021$, 
$\Omega_{c}h^{2}=0.111$, $\Omega_{\nu}h^{2}=0.006$. The latter is equivalent 
to a total neutrino mass $M_{\nu}=\sum_{i=1}^{3}m_{\nu,i}=$0.56eV.
Dark energy makes up the rest of the energy required for a spatially-flat universe. 
The Hubble constant, dark energy equation of state and helium fraction are, 
respectively, $H_{0}=70$ ${\rm km\ sec^{-1}Mpc^{-1}}$, $w=-1$ and $Y_{He}=0.24$. 
$h$ is the Hubble constant in 100 km/sec/Mpc units.
The optical depth to reionization and its redshift are $\tau_{re}=0.073$ and 
$z_{re}=12$. The normalization of the primordial power spectrum was set to 
$A_{s}=2.4\times 10^{-9}$ and its power law index is $n_{s}=0.947$. 

\section{Discussion}

CMB observations, especially those which will have the sensitivity and fidelity 
to detect the `forbidden' power spectra $C^{TB}$ and $C^{EB}$, will be able to set 
tight constraints on CB, a phenomena related to the bedrock of fundamental physics 
and its symmetries. Cosmology improves on terrestrial experiments in this context 
due to the long optical path lengths which enable the detection, in principle, of 
extremely small effects. However, since a statistically significant detection of 
CB heavily relies on measuring the TB and EB `smoking gun' correlations, and due 
to the fact that these involve the sub-$\mu K$ B-mode signal which is prone to numerous 
systematics, a credible detection of non-vanishing $\gamma$ should account for 
these possible sources of confusion. We saw that beam-rotation, differential 
ellipticity, as well as differential pointing, gain and beamwidth (to a lesser 
extent) can bias the inferred $\gamma$ but also change the uncertainty (in the case of 
differential pointing and ellipticity) due to their direct effect on the TB 
and EB correlations. While the differential pointing is very small due to its 
shallow l-dependence (compared to other systematics considered here) and 
the effect of beam ellipticity can be partially harnessed (Appendix A), it is 
the beam rotation effect which mainly contaminates CB. In particular, we have 
shown that CMB experiments which are optimized for inflationary and lensing 
science {\it may not} be adequate for the CB science that requires a much 
better control of beam rotation.
The nominal $\gamma$-detection with PLANCK and POLARBEAR is 3.8 and 
1.1 arcminutes, respectively. Beam rotation should therefore be controlled to 
0.75 and 0.21 arcminutes if a $5\sigma$ detection of CB is required. This systematic 
effect cannot be mitigated by scanning strategy. The effect of differential pointing 
(which is subdominant to the effect of differential ellipticity) can be further 
suppressed by scanning strategy mitigation [17]. The differential-ellipticity, 
together with the beam rotation effect, seem the most pernicious for TB spectra 
(Fig. 1).
We consistently assumed the worst-case-scenario as we did when we employed the 
constraints on $C^{BB}$ by assuming the tilt of the polarization direction to 
the ellipse major axis is $45^{\circ}$ [3] (see also Figure 2). However, an 
estimate of the real uncertainty on beam ellipticity (Appendix A) shows that, 
in practice, beam ellipticity should not be a major problem to TB measurements 
by PLANCK and POLARBEAR. The EB cross-correlations, however, should vanish if 
we take $\psi=45^{\circ}=\theta$. Nevertheless, for the purpose of illustration, 
and in the plots only, we assumed $\psi=22.5^{\circ}=\theta$ so as to maximize 
the EB correlations for the given beam parameters [3]. We further explored the 
effect of randomness of the angle $\psi$ in a multi-pixel elliptical-beam 
experiment (Appendix B) and found that our worst-case-scenario is actually 
representative of an average multi-pixel experiment. Similar considerations 
were also applied to beam rotation.

\section{Conclusion}

CB is an interesting ancillary science that future CMB polarimeters will target. 
We considered in detail two such experiments: PLANCK and POLARBEAR. Our past 
experience taught us that controlling beam systematics to a sufficiently high precision
as to aim at the B-mode detection (in case the tensor-to-scalar 
ratio is $r\approx 0.01$) will also automatically gurantee a high fidelity 
lensing-induced B-mode measurement. However, as we show here, this minimum 
requirement will in general not suffice for a credible CB detection via 
the `forbidden' TB and EB correlations. The likely reason for this is that while 
BB correlations are quadratic in the beam-imperfection parameters 
(e.g. $\propto e^{2}, \varepsilon^{2}$) the TB correlations are only linear 
in these {\it small} beam parameters (and the EB correlation is in general noisy). 
This implies that for a given ellipticity or beam rotation the fractional bias 
in the TB cross-correlations will be larger than the the corresponding fractional 
bias in the BB power spectrum by O(1/e) and O(1/$\varepsilon$), respectively. 

Whereas the effect of differential pointing on the TB and EB power spectra, 
we showed, is negligible, the effect of ellipticity is larger and can 
compromise the CB science if only the minimal requirement of the 
inflation/lensing science is satisfied. However, we demonstrated 
(Appendix A) that in practice the expected level of ellipticity 
uncertainty is relatively small (if the beam is standardly calibrated 
against a bright point source). The most pernicious effect is due to beam 
rotation which precisely mimics BC, has the same $l$-dependence, and cannot 
be mitigated by idealizing the scanning strategy. Our conclusion is that 
while TB and EB are unique indicators for new-physics in principle, they 
can in practice be excited by imperfect beams and in order to realize the 
promising potential of the high-sensitivity and fine-resolution PLANCK 
and POLARBEAR at CB detection beam rotation has to be controlled to the 
sub-arcminute level. Unless beam rotation is controlled to the arcminute level, 
a conservative approach, which does not use $C_{l}^{TB}$ and $C_{l}^{EB}$, may 
be more adequate for the CB science.

\section*{Acknowledgments}

We acknowledge the use of the publically available code by Lesgourgues, 
Perotto, Pastor \& Piat for the calculation of the noise in lensing 
reconstruction. We also acknowledge using CAMB for power spectra calculations. 
BK gratefully acknowledges support from NSF PECASE Award AST-0548262.

\begin{table}[c]
\begin{tabular}[c]{|c|c|c|c|}
\hline
~effect~&~parameter~&~$\Delta C_{l}^{TB}$~&~$\Delta C_{l}^{EB}$~\\
\hline
~gain~& $g$ & 0 & 0 \\
\hline
~monopole~& $\mu$ & 0 & 0\\
\hline
~pointing~& $\rho$ & $\frac{1}{2}[1+J_{0}(l\rho)]J_{2}(l\rho)s_{\theta}C_{l}^{T}$ & $s_{\theta}c_{\theta}J_{2}^{2}(l\rho)C_{l}^{T}$\\
~~&~~&~$+\frac{f_{3}}{2\pi}s_{\theta}J_{1}^{2}(l\rho)C_{l}^{T}$~&~~\\
\hline
~quadrupole~& $e$ & $-I_{0}(z)I_{1}(z)s_{\psi}C_{l}^{T}$ & 
$I_{1}^{2}(z)s_{\psi}c_{\psi}C_{l}^{T}$ \\
\hline
~rotation~& $\varepsilon $ & $\sin(2\varepsilon) C_{l}^{TE}$ & $\frac{1}{2}\sin(4\varepsilon) C_{l}^{E}$\\
\hline
\end{tabular}
\caption{The contribution of the systematic effects to 
the power spectra $C_{l}^{TB}$, $C_{l}^{EB}$
assuming the underlying sky is not polarized (except 
for the {\it rotation} signal when we assume E-, and B-mode polarization are present) 
and ideal sky scanning. The definitions of $s_{\theta}$ and $s_{\psi}$ as well as $f_{3}$ 
can be found in [17] where similar expressions for $C_{l}^{TE}$, $C_{l}^{EE}$ and $C_{l}^{BB}$ 
are also found and which we employed in this work.}
\end{table}

\begin{table}[c]
\begin{tabular}[c]{|c|c|c|c|}
\hline
~Experiment~&~$\sigma_{\gamma}^{{\rm nominal}}[{\rm deg}]$~&~{\rm parameter}~&~$\Delta\gamma$[{\rm deg}]\\
\hline
~POLARBEAR~&~0.0179~&~$\rho$~&~0.0094~\\
\cline{3-4}
~~&~~&~$e$~&~56.07~\\
\hline
~PLANCK~&~0.0641~&~$\rho$~&~0.0060~\\
\cline{3-4}
~~&~~&~$e$~&~93.9~\\
\hline
\end{tabular}
\caption{Estimated bias in the inferred rotation angle assuming POLARBEAR and PLANCK 
were optimized to detect the inflationary B-mode signal associated with r=0.01 ([3]). 
The bias induced by differential ellipticity is much larger than the naive uncertainty, 
highlighting the need in careful data analysis when interpretation of polarization 
plane rotation is required. The relatively small bias induced by the pointing is 
negligible compared to the ellipticity-induced bias and this is 
consistent with the findings of [3] and explained by the steep l-dependence of the ellipticity 
systematics.}
\end{table}

\begin{figure*}
\begin{tabular}{cc}
\includegraphics[width=\columnwidth]{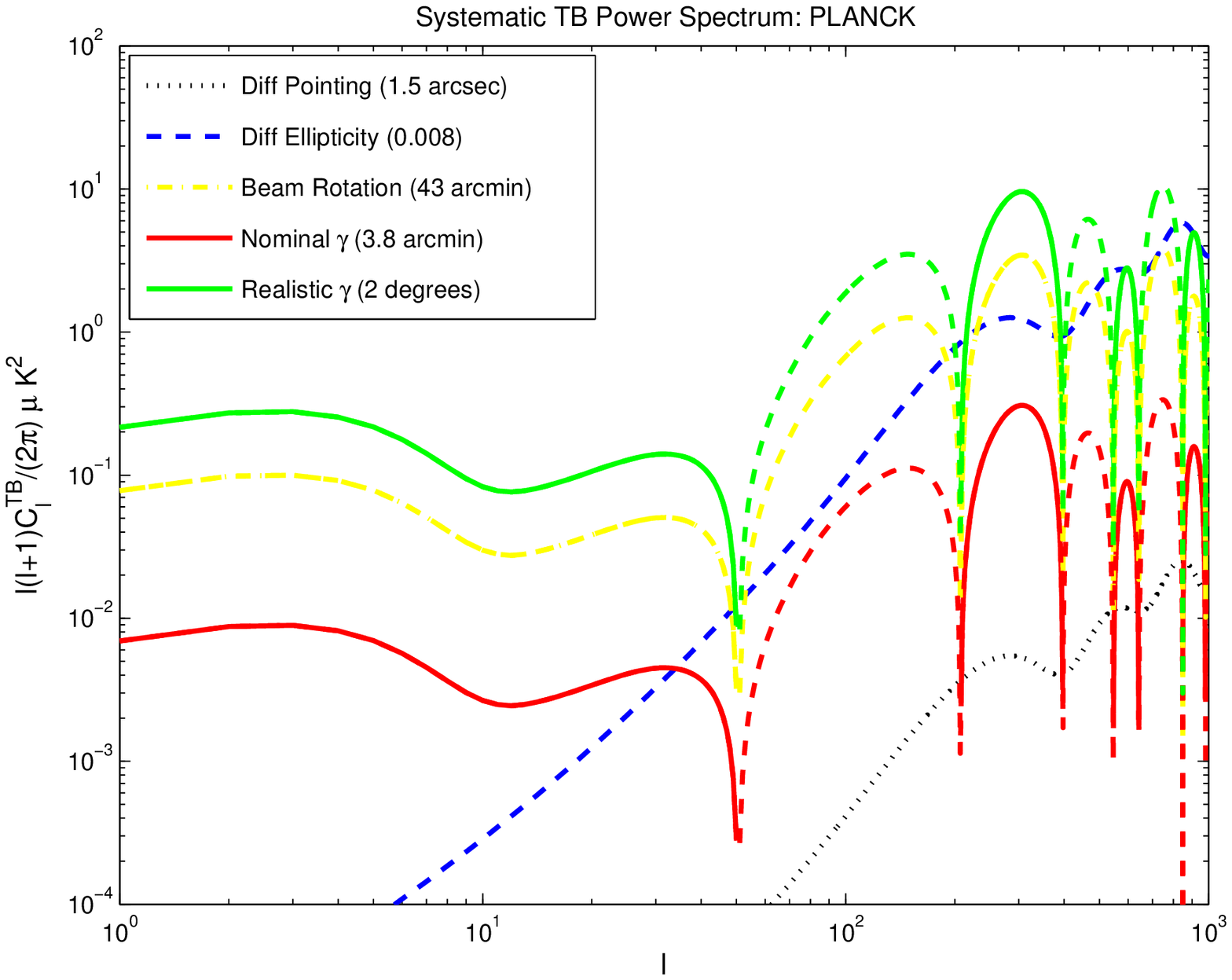} &
\includegraphics[width=\columnwidth]{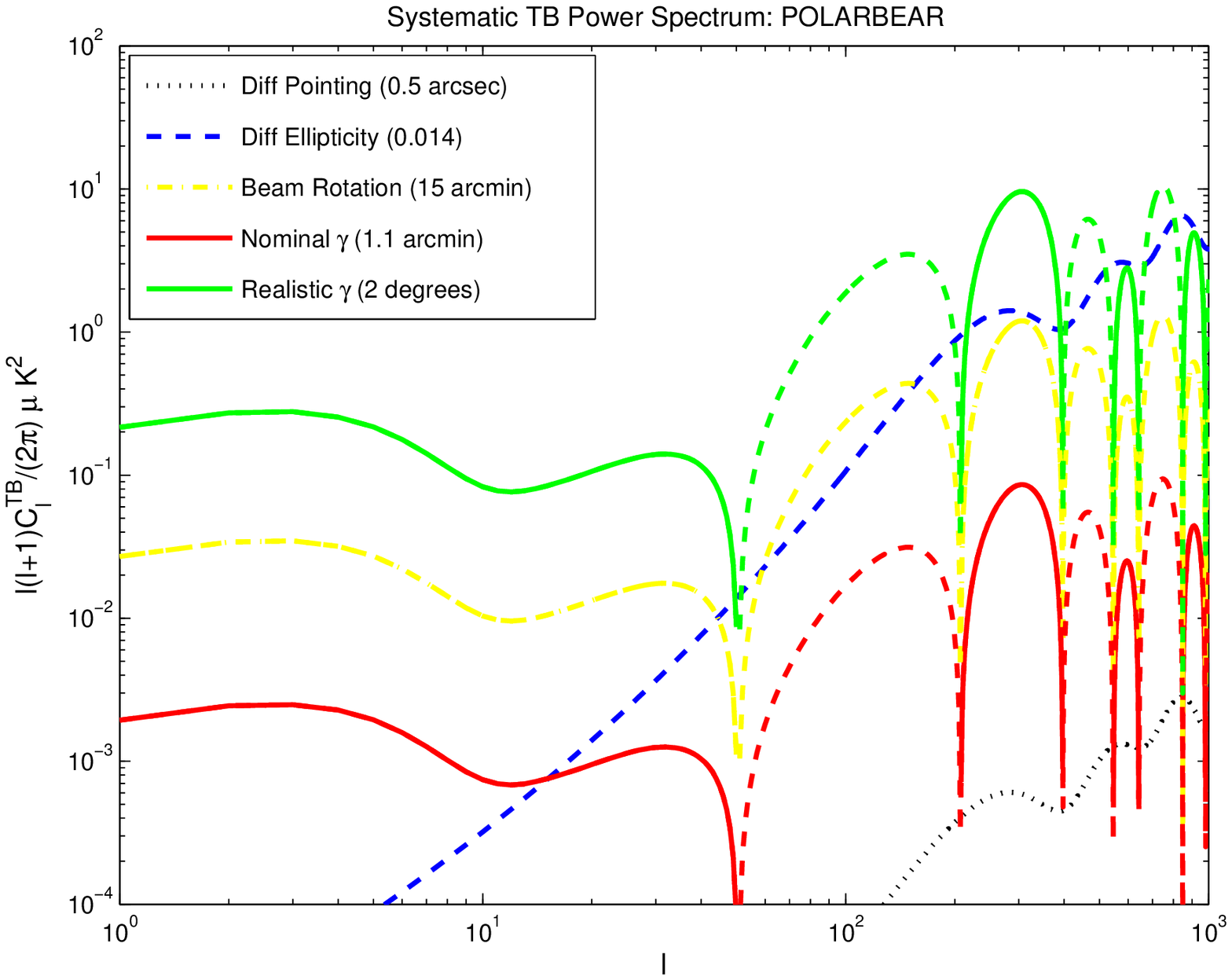} \\
\includegraphics[width=\columnwidth]{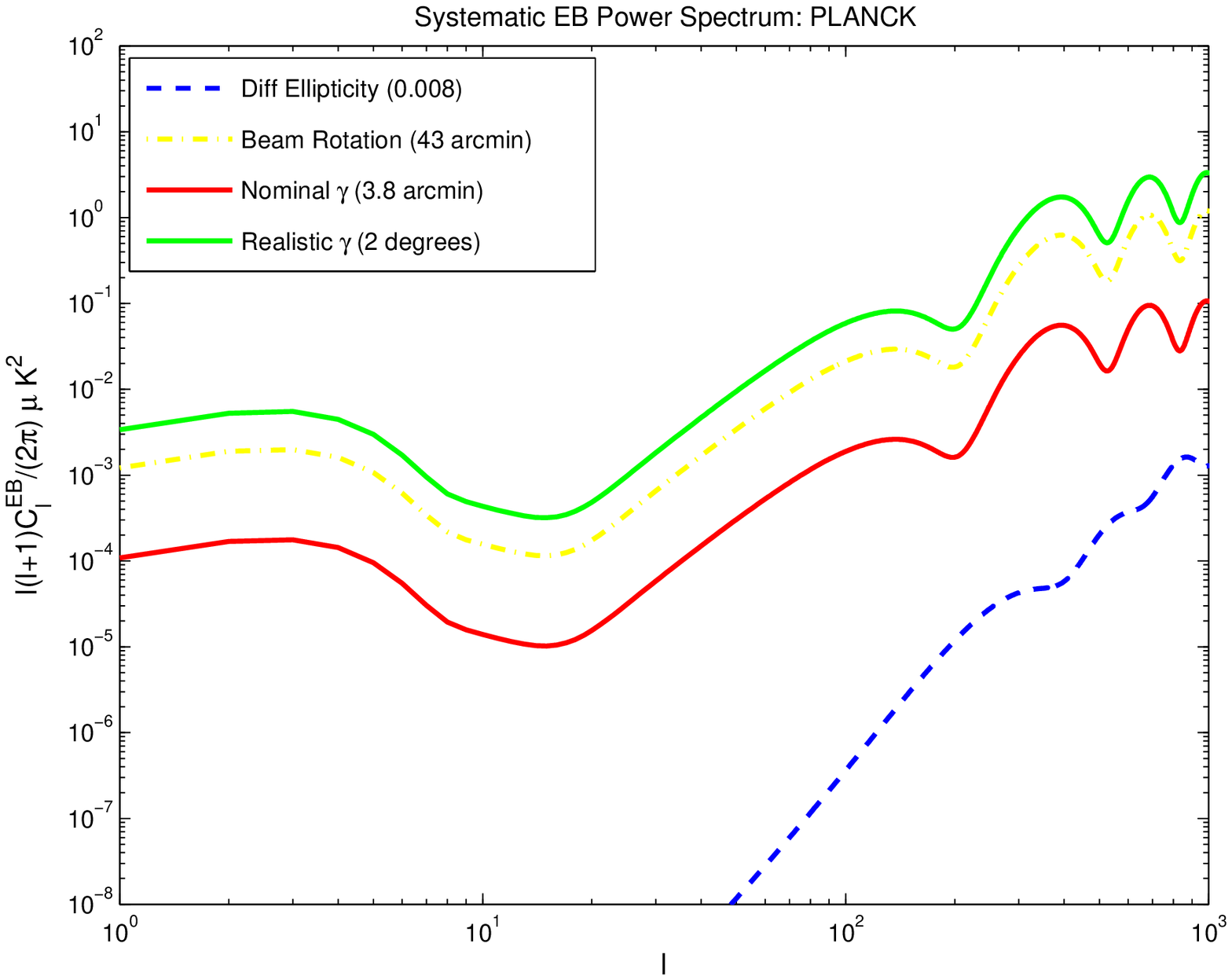} &
\includegraphics[width=\columnwidth]{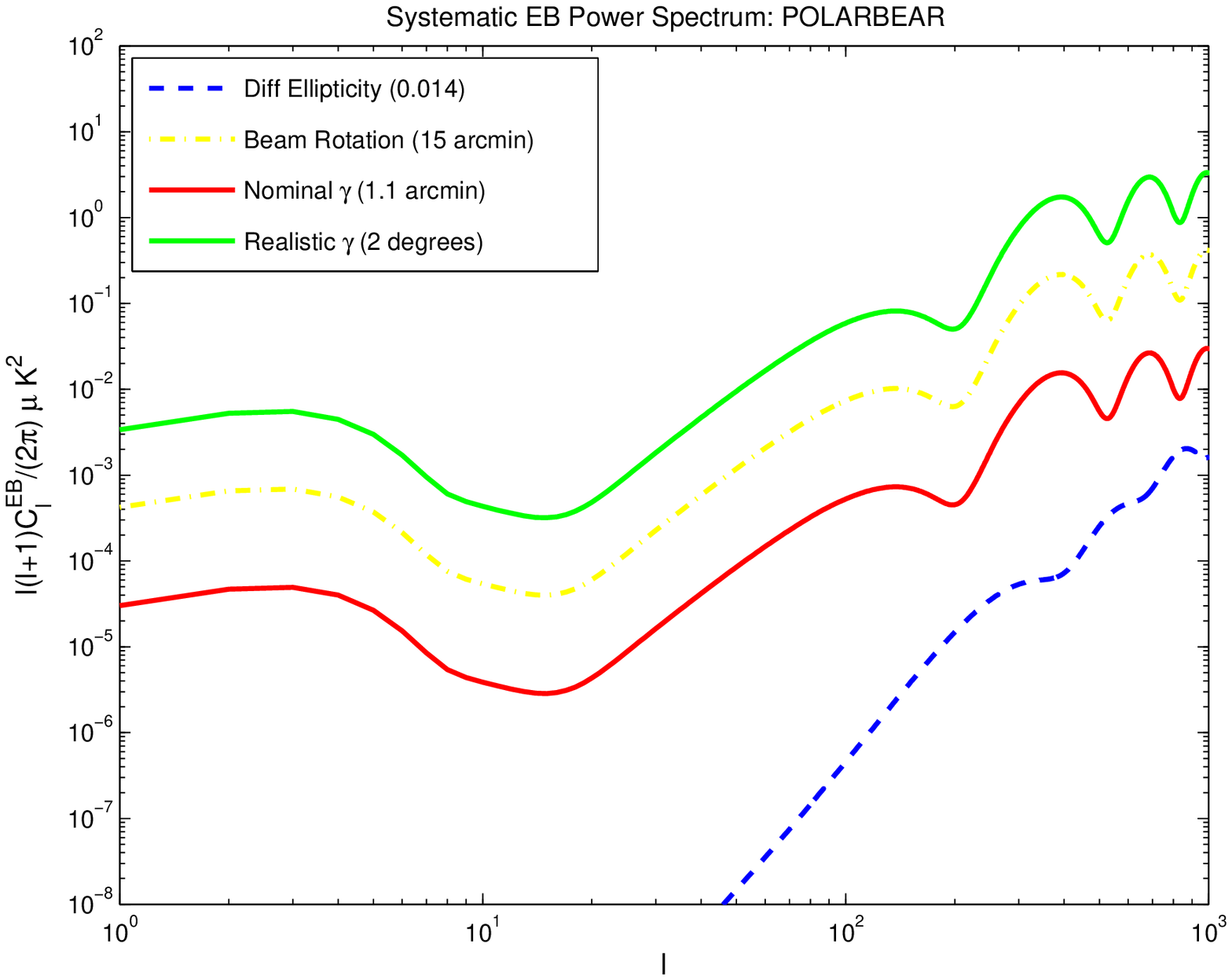}
\end{tabular}
\caption{TB and EB systematics compared to the nominal CB signals 
(i.e. those we expect to obtain in the absence of beam systematics):
black (dotted), blue (dashed) and yellow (dotted-dashed) curves refer to the systematics induced by 
differential pointing and ellipticity and by beam rotation, respectively. 
Red and green (continuous) curves represent the cross-correlations in the nominal and 
realistic CB cases, respectively. `Realistic' is defined as the CB that 
would be roughly three times larger than the largest systematic. 
Dashed-dotted curves represent the absolute values of negative T-B 
cross-correlations in certain $l$-ranges. 
All signals are calculated based on the assumption 
that the beam systematics of PLANCK and POLARBEAR are optimized for 
the detection of the inflationary signal at the $r=0.01$ 
level [3]. These are used in the analytic expressions of  
[17] for $C_{l}^{TB}$ and $C_{l}^{EB}$. It is clear from the upper 
panel that the main contaminant is due to differential ellipticity and 
beam rotation.}
\end{figure*}

\subsection{Appendix A: Forecasted Beam Ellipticity}

It is instructive to estimate the expected uncertainty of beam ellipticity which 
may result from a standard beam-calibration procedure with 
a nearly black-body point-source such as Jupiter ($T_{p}\approx 150K$, 
$\theta_{p}\approx 0.5$ arcmin). By Wiener filtering [4] a map of the 
expected signal one expects to recover the source image with a signal-to-noise level 
\begin{eqnarray}
\left(\frac{S}{N}\right)^{2}
=\int\frac{|\tilde{S}({\bf l})|^{2}}{P({\bf l})}\frac{d^{2}{\bf l}}{(2\pi)^{2}}.\nonumber
\end{eqnarray}
\begin{flushright}
(A.1)
\end{flushright}
where $\tilde{S}({\bf l})$ is the Fourier transform of the point source and 
$P(l)$ is the instrumental noise, i.e. the $(S/N)^{2}$ is the ratio of the 
signal and noise power-spectra integrated over all accessible multipoles in 
the experiment in question. The higher S/N the smaller are the uncertainties 
in the recovered beam parameters.
The Fourier transform of the convolved calibration source is
\begin{eqnarray}
\tilde{T}_{p}^{\rm obs}=\tilde{T}_{p}
e^{-\frac{1}{2}l_{x}^{2}\sigma_{x}^{2}
-\frac{1}{2}l_{y}^{2}\sigma_{y}^{2}-i{\bf l}\cdot\rho}\nonumber
\end{eqnarray}
\begin{flushright}
(A.2)
\end{flushright}
where we assume an elliptical gaussian beam with principal axes 
$\sigma_{x}$ and $\sigma_{y}$, and pointing ${\bf \rho}$. This results in 
\begin{eqnarray}
\left(\frac{S}{N}\right)^{2}=\frac{2\ln(2)}
{\pi (1-e^{2})}\left(\frac{T_{p}}{\Delta_{b}}\right)^{2}\left(\frac{\theta_{p}}
{\theta_{b}}\right)^{4}\eta^{2}f_{t}\nonumber
\end{eqnarray}
\begin{flushright}
(A.3)
\end{flushright}
where $\eta$ is an experiment-specific optical-efficiency parameter ($\le 1$) 
and $f_{t}$ is the fraction of observation time dedicated to beam calibration 
(we adopt the value 1/30). $\Delta_{b}$ is the instrumental temperature equivalent 
noise, $\theta_{b}$ is the beamsize and $e$ is its ellipticity. 
Since the pointing merely adds a phase to the beam function it drops from 
the expression for S/N. Similarly, S/N is also independent of the beam 
rotation angle since temperature measurements are insensitive to $\varepsilon$. 
Calibrating the beam rotation with a polarized source will not work here since 
by analogy to Eq. (A.1) the rotation only adds a phase.
Therefore, the following procedure, which is based on S/N considerations, 
will be used to determine the uncertainty of $e$ only.
To estimate these uncertainties we require that varying the beam parameters 
results in signal changes smaller than the noise. That is, we allow the 
uncertainty in the beam's ellipticity to grow from 0 until the $(S/N)^{2}$ changes by unity 
\begin{eqnarray}
(S/N)^{2}\rightarrow (S/N)^{2}+1.\nonumber
\end{eqnarray}
\begin{flushright}
(A.4)
\end{flushright}
This condition readily yields the uncertainty in beam parameters
\begin{eqnarray}
\Delta (e^{2})=[S\eta/N]^{-2}f_{t}^{-1}.\nonumber
\end{eqnarray}
\begin{flushright}
(A.5)
\end{flushright}
Assuming Jupiter is a $150K$ black body with a characteristic scale of 
$\theta_{p}=0.5'$ , typical beamsizes 
$\theta_{b}$ are 5' (4') and instrumental noise $\Delta_{b}$ $1.7\mu K$ 
($10\mu K$) for PLANCK (POLARBEAR) as 
well as 5\% optical efficiency $\eta$ 
we obtain the following estimates on expected uncertainty in beam ellipticity 
$e=1.1\times 10^{-3}$ (PLANCK) and $e=1.2\times 10^{-4}$ (POLARBEAR). 
Using the scaling of ellipticity-induced 
TB and EB correlations ($\propto e$ and $\propto e^{2}$, respectively), 
it is clear that based on this estimate beam ellipticity is unlikely 
to be the dominant contaminant of the EB correlations but the TB 
correlations will still be somewhat biased even for the optimal ellipticity 
we find here (O($10^{-3}$) for PLANCK and O($10^{-4}$) for POLARBEAR).

\subsection{Appendix B: Focal Plane Considerations}

Previous works [1, 2, 3, 17] assumed, for simplicity, single-pixel experiments. 
However, in practice the beam-pairs are scattered in the focal plane with 
different polarization orientations and other beam parameters such as gain, 
beamwidth, ellipticity, etc. We briefly discuss how the random orientation 
of beam polarization only mildly suppress spurious B-mode polarization.
A schematic focal plane is depicted in Fig.2. 
\begin{figure*}[hb]
\begin{tabular}{cc}
\includegraphics[width=10cm]{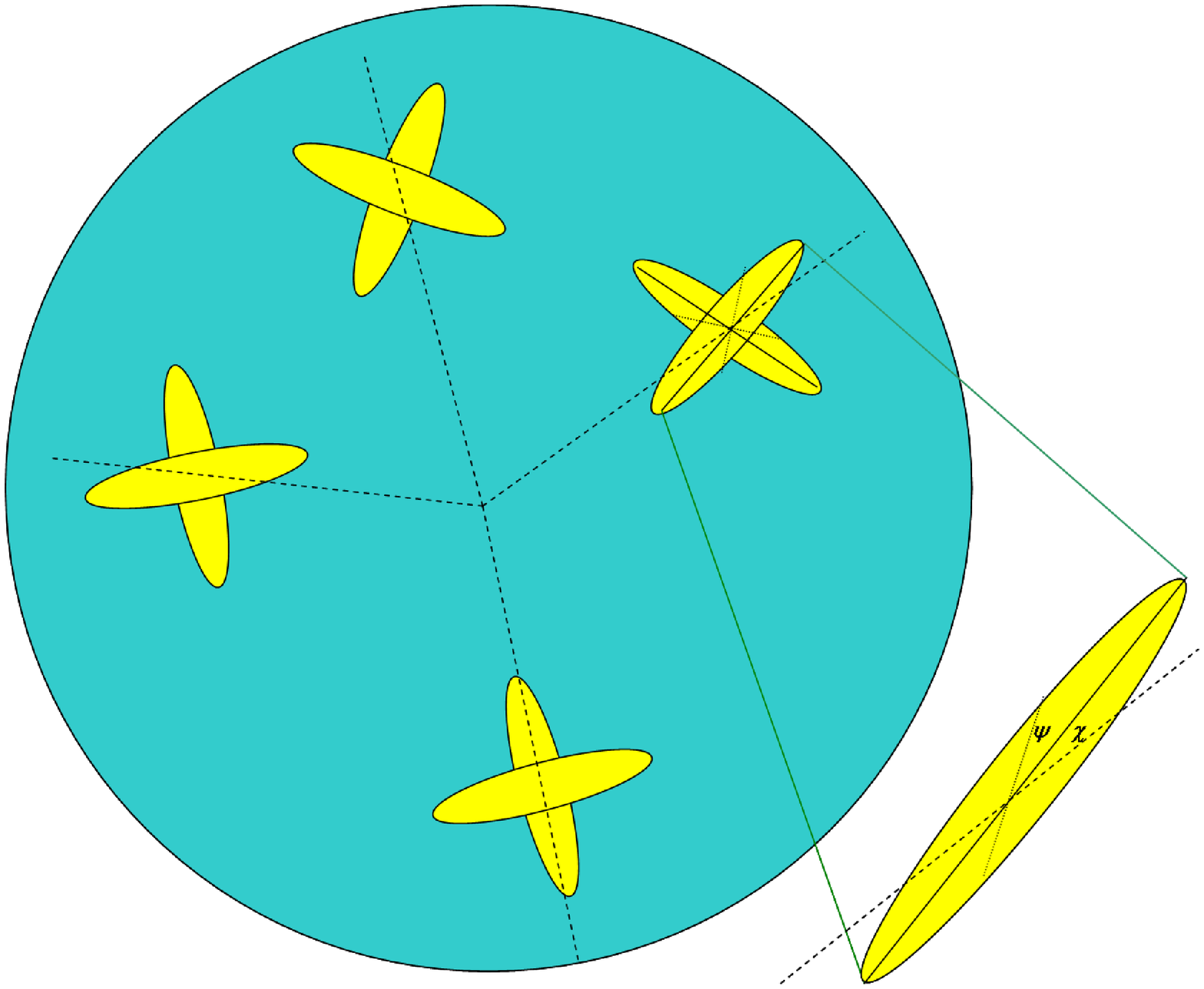}
\end{tabular}
\caption{Layout of typical beam pairs in the focal plane: dashed lines 
represent radius-vectors from the center of the focal plane through the 
center of each pair. Continuous lines represent the ellipse major 
axes and the angle the major axis makes with the radius-vector is $\chi$. 
The polarization axis (dotted line) makes an angle $\psi$ with the major axis. 
In [17] and [3] it was assumed that $\chi=0$, therefore every $\psi$ 
in our analytic expressions (Table 1) should be replaced with $\psi+\chi$.}
\end{figure*}  
We assume all beams have the same ellipticity for the purpose 
of illustration but they differ by their orientations with respect to 
the radius-vector which connects the center of the focal plane to the 
beam center (denoted $\chi$) and the polarization axis makes an angle 
$\psi$ with the ellipse major axis. We further assume that the major 
axes of the beams are nearly orthogonal ($\chi_{1}-\chi_{2}\approx 90^{\circ}$) 
and similarly their respective polarization directions satisfy 
($\psi_{1}-\psi_{2}\approx 90^{\circ}$), where the 
subscripts stand for `beam 1' and `beam 2' of each pair, respectively. 
We calculate the covariance of the individual systematic power spectra
\begin{eqnarray}
cov(C_{l}^{Y},C_{l'}^{Y})\equiv\delta_{l,l'}[\langle (C_{l}^{Y})^{2}\rangle - 
(\langle C_{l}^{Y}\rangle)^{2}]\nonumber
\end{eqnarray}
\begin{flushright}
(B.1)
\end{flushright}
where $Y$ stands for either $BB$, $TB$ or $EB$ and the angular brackets denote 
averages over the angles $\psi$ and $\chi$. 
Since the number of pairs is finite, all the estimates we obtain here for 
the suppression of systematics by assuming infinitely many detector pairs constitute 
only upper limits. Nevertheless, we show that even if the number of detectors is assumed 
infinite, the gain in systematic suppression is insignificant. Taking the ratio 
of the covariance to the worst-case $C_{l}^{B}$, $C_{l}^{TB}$ and $C_{l}^{EB}$ 
($\psi+\chi=45^{\circ}$, $45^{\circ}$ and $22.5^{\circ}$, respectively) it is 
straight forward to show that with an infinite number of pairs the 
systematic $C_{l}^{B}$, $C_{l}^{TB}$ and $C_{l}^{EB}$ drop 
to 71\%, 71\% and 50\% of their worst-case values - an insignificantly 
small change to our conclusions. We can apply a similar calculation to beam rotation.
in this case averaging is carried out over the angle $\varepsilon$ (random orientations 
of the pair subject to the constraint that the polarization directions of the two polarimeters 
remain orthogonal). Here we obtain that $C_{l}^{B}$, $C_{l}^{TB}$ and $C_{l}^{EB}$ are suppressed 
to 35\%, 71\% and 50\% of their worst-case values. Again, this will not quanlitatively change our 
conclusions.


\begin{thebibliography}{99}
\bibitem{1} Hu, W., Hedman, M.~M., \& Zaldarriaga, M.\ 2003, PRD, 67, 043004
\bibitem{2} O'Dea, D., Challinor, A., \& Johnson, B.~R.\ 2007, MNRAS, 376, 1767
\bibitem{3} Miller, N.~J., Shimon, M., \& Keating, B.~G.\ 2008, PRD, 79, 063008 
\bibitem{4} Smith, K.~M., et al.\ 2008, arXiv:0811.3916 
\bibitem{5} Carroll, S.~M., Field, G.~B., \& Jackiw, R.\ 1990, PRD, 41, 1231
\bibitem{6} Carroll, S.~M., \& Field, G.~B.\ 1997, PRL, 79, 2394
\bibitem{7} Carroll, S.~M.\ 1998, PRL, 81, 3067 
\bibitem{8} Lue, A., Wang, L., \& Kamionkowski, M.\ 1999, PRL, 83, 1506
\bibitem{9} Alexander, S.~H., Peskin, M.~E., \& Sheikh-Jabbari, M.~M.\ 2006, PRL, 96, 081301
\bibitem{10} Alexander, S., \& Martin, J.\ 2005, PRD, 71, 063526 
\bibitem{11} Feng, B., Li, M., Xia, J.-Q., Chen, X., \& Zhang, X.\ 2006, PRL, 96, 221302
\bibitem{12} Liu, G.-C., Lee, S., \& Ng, K.-W.\ 2006, PRL, 97, 161303
\bibitem{13} Cabella, P., Natoli, P., \& Silk, J.\ 2007, PRD, 76, 123014 
\bibitem{14} Xia, J.-Q., Li, H., Wang, X., \& Zhang, X.\ 2008, A\&A, 483, 715
\bibitem{15} Komatsu, E., et al.\ 2008, ArXiv e-prints, 803, arXiv:0803.0547 
\bibitem{16} Wu, E.~Y.~S.~, et al.\ 2008, arXiv:0811.0618
\bibitem{17} Shimon, M., Keating, B., Ponthieu, N., \& Hivon, E.\ 2008, PRD, 77, 083003 
\bibitem{18} Kamionkowski, M.\ 2008, arXiv:0810.1286 

\end{thebibliography}
\end{document}